\documentclass[twocolumn,showpacs,aps]{revtex4}

\usepackage{graphicx}
\usepackage{dcolumn}
\usepackage{bm}

\begin{document}

\title{$\mathbf{^{25}Mg}$ NMR  study of the $\mathbf{MgB_2}$
superconductor}
\author{M.~Mali}
\email{mali@physik.unizh.ch}
\author{J.~Roos, A.~Shengelaya and H.~Keller}
\affiliation{Physik--Institut, Universit\"{a}t Z\"{u}rich, CH--8057 Z\"{u}rich,
         Switzerland}
\author{K. Conder}
\affiliation{Laboratory for Neutron Scattering, ETH Z\"urich and PSI Villigen,
CH--5232 Villigen PSI, Switzerland}
\begin{abstract}
$\mathrm{^{25}Mg}$ NMR spectra and nuclear spin-lattice relaxation 
time, $T_1$, have been measured in polycrystalline $\mathrm{MgB_2}$ 
with a superconducting transition temperature $T_{c}$ = 39 K in zero 
magnetic field.  From the first order and second order quadrupole 
perturbed NMR spectrum a quadrupole coupling frequency $\nu_{Q}$ = 
222.0(1.5) kHz is obtained.  $T_1T = 1090(50)$ sK and Knight shift $K_c 
= 242(4)$ ppm are temperature independent in the normal conducting 
phase.  The $\mathrm{^{25}Mg}$ Korringa ratio equals to 0.95 which is 
very close to the ideal value of unity for s-electrons.  The 
comparison of the experimental $\nu_{Q}$, $T_1T$, and $K_c$ with the 
corresponding values obtained by LDA calculations shows an 
excellent agreement for all three quantities.
 \end{abstract}
 \maketitle
The recent discovery of superconductivity in $\mathrm{MgB_2}$ with
remarkably high $T_{c}$ = 39 K \cite{Nagamatsu} has attracted much
attention. In particular the observation of a sizeable boron
\cite{Budko} isotope effect strongly suggests that this simple
layered intermetallic compound belongs to the conventional family
of phonon mediated BCS superconductors.  The relevant
electron-phonon coupling constant is proportional to the density
of states (DOS) at the Fermi level. So it is important to have
experimental data of this quantity in $\mathrm{MgB_2}$.

Nuclear magnetic resonance (NMR) in metals probes the DOS at the Fermi 
level.  The measured quantities, the spin lattice relaxation rate, 
1/$T_1$, and the Knight shift, $K$, are related to the electron spin 
susceptibility ($\chi(\mathbf{q},\omega)$) of electrons close to the Fermi 
level, specifically 
$1/(T_1T) \propto \lim_{\omega\to0} 
\sum_{\mathbf{q}} \mathrm{Im\chi(\mathbf{q},\omega)/\omega}$, 
and $K \propto \mathrm{Re\chi(0,0)}$.  
In case the atomic site symmetry in the crystal structure is less than 
cubic and the atom has a nuclear quadrupole moment as it applies for 
Mg and B in $\mathrm{MgB_2}$, quadrupolar disturbed NMR delivers in 
addition valuable information on the electric field gradient (EFG) at 
the specific nuclear site.  Unlike 1/$T_1$ and $K$ the EFG is determined 
by the distribution of all charges.  Up to date only measurements of 
1/$T_1$, $K$, and of the quadrupole coupling to the EFG, expressed 
by quadrupole coupling frequency $\nu_{Q}$, 
of $\mathrm{^{11}B}$ were performed 
\cite{Kotegawa,Gerashenko,Jung,Tou}.  While there is reasonable 
agreement among $\mathrm{^{11}B}$ $\nu_{Q}$'s and relaxation rates, 
there is a considerable controversy concerning the $\mathrm{^{11}B}$ 
Knight shifts.  While some authors \cite{Gerashenko} report in the 
normal conducting state a small temperature independent isotropic 
shift of 175 ppm, others \cite{Jung} report a smaller and even 
temperature dependent shift of approximately 60 ppm, which they 
attribute to the Fermi-contact interaction.  Finally there is also a 
report \cite{Tou} of a negative $\mathrm{^{11}B}$ shift of mere 5 ppm 
attributed to the core polarization.  These large discrepancies most 
likely come from the choice of different materials as references for 
the Knight shift, as well as from difficulties to measure the small 
$\mathrm{^{11}B}$ Knight shift of a broad and in addition 
quadrupolarly shifted $\mathrm{^{11}B}$ central line powder spectrum.

As far as we know, there are no experimental data concerning the
NMR quantities at the Mg site in $\mathrm{MgB_2}$, although
theoretical predictions based on {\it ab initio} local density
approximation (LDA) calculations exist for Mg Knight shift,
1/$T_1$ and EFG \cite{Pavarini,Tsvyashchenko}. To some degree the
lack of experimental Mg NMR data is understandable considering the
fact that the only NMR active $\mathrm{^{25}Mg}$ isotope has a
small magnetic moment and low natural abundance.  Consequently the
$\mathrm{^{25}Mg}$ NMR signals are weak even in high magnetic
fields.  Their detection often demands prohibitively long
signal accumulation times.  Despite this handicap we were able to
measure NMR of the naturally abundant $\mathrm{^{25}Mg}$ in
$\mathrm{MgB_2}$.  Here we report the temperature dependence of
Knight shift, spin-lattice relaxation time, and quadrupole
coupling frequency of $\mathrm{^{25}Mg}$ in $\mathrm{MgB_2}$.  We
compare the experimental results with the theoretical predictions
and we find for all three Mg NMR quantities an excellent
agreement. Since all calculations were done by state of the art
LDA methods the good agreement between calculated and measured NMR
quantities confirms the LDA as a good approximation for $\mathrm{MgB_2}$.

\begin{figure}[h]
   \centering
   \includegraphics[width=\linewidth]{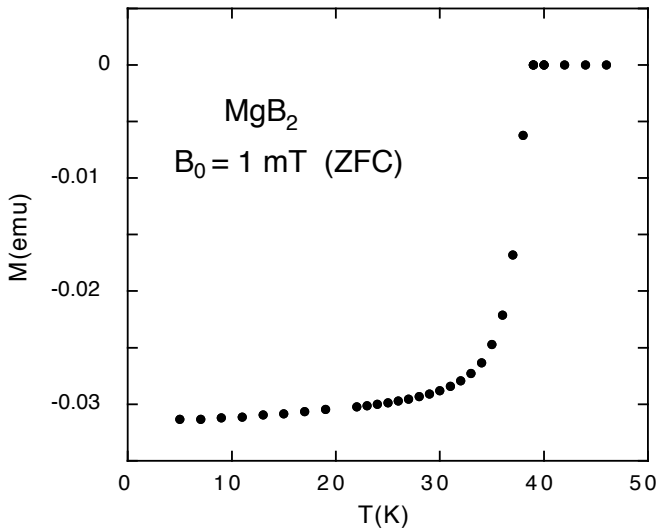} \vskip12pt
    \caption{Temperature dependence of DC magnetization of $\mathrm{
    MgB_2}$ powder in a magnetic field of 1 mT (ZFC).}
    \label{fig1}
\end{figure}

The $\mathrm{MgB_2}$ sample was prepared using stoichiometric amounts 
of magnesium and boron (99$\%$ and 99.99$\%$, respectively) in a form 
of powder.  Both components were thoroughly mixed and pressed into 
pellets.  These were placed in a tantalum crucible equipped with a 
non-vacuum tight cover.  The crucible was then sealed under vacuum in a 
quartz ampoule.  The sample was synthesized during heating at 600, 800 
and 900 $^{\circ}$C for one hour at each temperature.  X-ray diffraction 
(XRD) has revealed only small amount of $\mathrm{MgO}$ as an impurity 
phase. DC magnetization measurements (Fig. \ref{fig1}) in a magnetic
field of 1 mT (ZFC) yield a transition temperature of 39 K.

\begin{figure}[h]
   \centering
   \includegraphics[width=\linewidth]{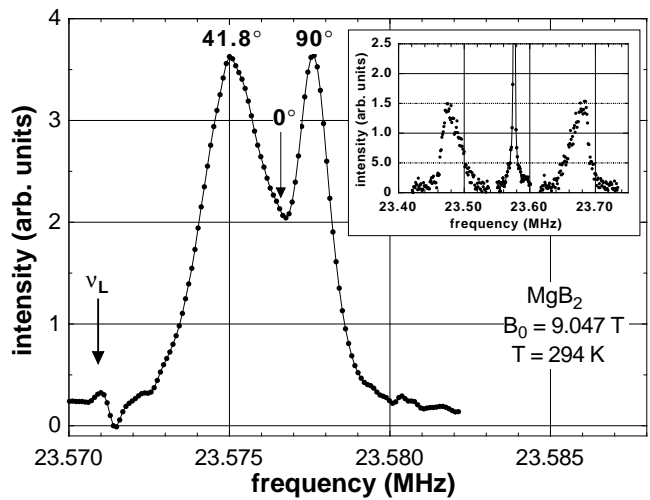} \vskip12pt
    \caption{$\mathrm{^{25}Mg}$ central line of $\mathrm{MgB_2}$ 
    powder spectrum measured at T = 294 K in a magnetic field $B_0 = 
    9.047$ T. Insert: $\mathrm{^{25}Mg}$ first satellites' singularities of $\mathrm{MgB_2}$ 
    powder spectrum measured at T = 294 K in a magnetic field $B_0 = 9.047$ T.}
    \label{fig2}
\end{figure}

\begin{figure}[h]
   \centering
   \includegraphics[width=\linewidth]{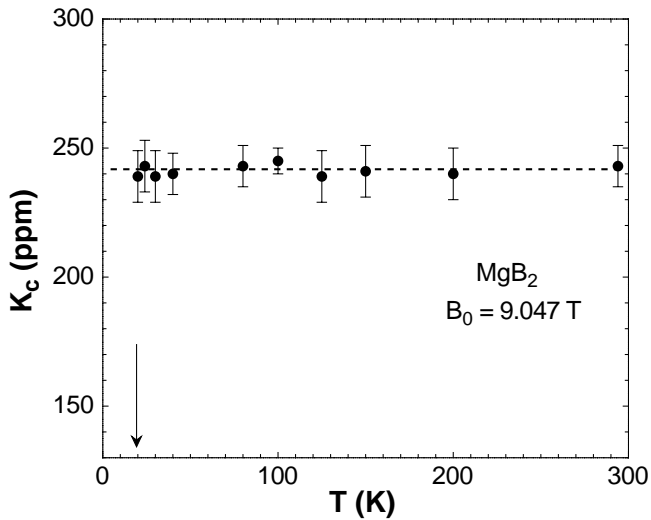} \vskip12pt
    \caption{Temperature dependence of the $\mathrm{^{25}Mg}$ Knight 
    shift, $K_c$, in normal conducting $\mathrm{MgB_2}$ powder grains 
    with c $\parallel B_0$. The shift is with respect to the $\mathrm{
    ^{25}Mg}$ resonance in a water solution of $\mathrm{MgCl_2}$ at 
    room temperature. The dashed line represents the weighted average 
    of $K_c$. The arrow in the figure marks the transition temperature 
    of $\mathrm{MgB_2}$ powder grains with c $\perp B_0 = 9.047$ T .}
    \label{fig3}
\end{figure}

\begin{figure}[h]
   \centering
   \includegraphics[width=\linewidth]{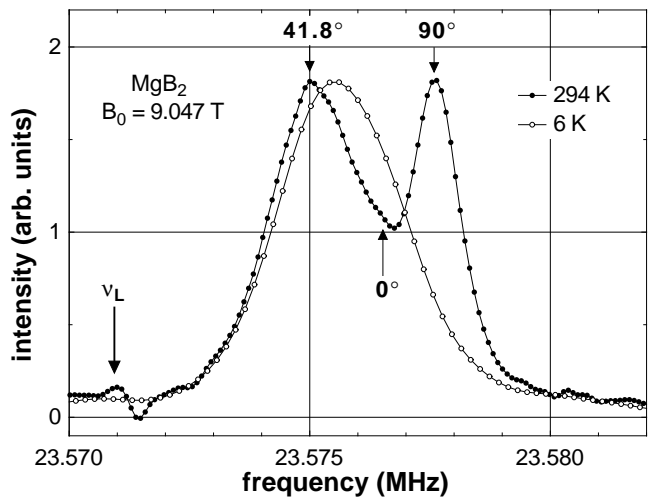} \vskip12pt
    \caption{$\mathrm{^{25}Mg}$ central lines of $\mathrm{MgB_2}$ 
    powder spectra measured at T = 6 K (empty circles) and at T = 
    294 K (filled circles) in a magnetic field $B_0 = 9.047$ T. 
    The intensities of the two spectra are scaled to equal height for 
    easier comparison .}
    \label{fig4}
\end{figure}

Before we proceed let us first present a few NMR relevant 
characteristics of the $\mathrm{^{25}Mg}$ isotope.  The isotope's 
natural abundance is 10.0$\%$ and its nucleus has spin 5/2, a 
gyromagnetic ratio $\gamma = 1.637\times10^{7} 
\mathrm{rad\,s^{-1}T^{-1}}$, and a quadrupole moment $Q = 
0.1994\times10^{-28} \mathrm{m^{2}}$ \cite{web elements}.  $\mathrm{
^{25}Mg}$ NMR measurements were performed on a 
polycrystalline $\mathrm{MgB_2}$ powder sample with a pulsed 
Fourier-transform spectrometer at an external magnetic field $B_0 = 
9.047$ T.  Fig. \ref{fig2} presents the $\mathrm{^{25}Mg}$ central 
line powder spectrum gained by Fourier transform of the free induction 
decay signal induced by a $\mathrm{7\mu s}$ $\pi/2$-pulse.  
The spectrum shows the typical powder pattern of the central $\mathrm{+1/2 
\leftrightarrow -1/2}$ transition line of a spin 5/2 system whose $\pm 
1/2$ levels are disturbed in second order by quadrupole coupling to an 
axially symmetric EFG.  There are two peaks in the spectrum with an 
indication of a washed out step left and close to the minimum between the 
peaks.  The three 
features from left to right (see Fig. \ref{fig2}) are the 
extrema singularities that occur at three powder grain orientations 
having either $\theta = 41.8^{\circ}, 0^{\circ}$, or $90^{\circ}$.  
Here $\theta$ represents the angle between the largest principal 
component, $V_{zz}$, of the EFG tensor and the magnetic field 
direction.  Due to the site symmetry the EFG tensor at Mg and B sites 
is axially symmetric along the c-axis. Therefore its $V_{zz}$ points 
into the c-axis direction.  From the distance between the two peaks in 
the spectrum and by help of the second order quadrupole effect 
expression \cite{Abragam}
\begin{displaymath}
   \Delta\nu = 25 [I(I+1) - 3/4] \frac{\nu_{Q}^{2}}{144 \nu_{L}} =
   \frac{25 \nu_{Q}^{2}}{18 \nu_{L}},
\end{displaymath}
we can calculate $\nu_{Q}$, defined as
\begin{displaymath}
   \nu_{Q} \equiv \frac{3eQ V_{zz}}{2I(2I-1) h} = \frac{3eQ V_{zz}}{20 h}.
\end{displaymath}
Having $\mathrm{\Delta\nu = 2.85(15)}$ kHz and Larmor frequency 
$\mathrm{\nu_{L} = 23.570870}$ MHz, as determined by $\mathrm{
^{25}Mg}$ NMR from $\mathrm{MgCl_{2}}$ in water solution, one gets 
$\mathrm{\nu_{Q} = 220(6)}$ kHz. In addition we were able to 
see the singularities of the first satellites in the satellite powder 
spectrum presented in the insert of  Fig. \ref{fig2}.  From the separation of the two first 
satellite singularities, that are positioned symmetrically with respect 
to the central line , one gets immediately $\nu_{Q}$. For an 
axially symmetric EFG  $\nu_{Q}$ is simply equal to this separation.  The gained 
$\mathrm{\nu_{Q} = 222.0(1.5)}$ kHz agrees very well with the above 
result.  The second result, however, is always more reliable 
especially in case when the additional Knight shift of the spectrum is 
anisotropic.  Since for $\theta = 0^{\circ}$ there is no quadrupole 
shift of the central line the $\theta = 0^{\circ}$ step in the 
spectrum can be used to determine the Knight shift, $K_{c}$, of the 
crystallites whose c-axis is parallel to the magnetic field.  From the 
shift of the step with respect to $\nu_{L}$ we get then $K_{c} = 
242(4)$ ppm.  Further we noticed that the ratio of the distances of 
the $\theta = 41.8^{\circ}$ and $90^{\circ}$ singularities with 
respect to the step is somewhat less than the value 16/9 
theoretically expected for an isotropic magnetic shift of the central line powder 
spectrum.  This allows the conclusion that the Knight 
shift has to be slightly anisotropic.  The Knight shift, $K_{ab}$, for 
crystallites whose c-axis is perpendicular to $B_0$ must be a bit 
larger than $K_{c}$.  We estimate the difference between $K_{ab}$ and 
$K_{c}$ to be about +10 ppm.  In the normal conducting phase from 294 K down 
to 19 K we observe that the Knight shift (see Fig. \ref{fig3}) 
and $\nu_{Q}$ = 222.0(1.5) kHz remain in error bar limits constant.  
Such a behavior was also observed for 
$\mathrm{^{11}B}$ Knight shift \cite{Gerashenko} and $\nu_{Q}$ 
\cite{Gerashenko,Jung}.  At 19 K the shape of the Mg central line 
spectrum changes drastically .  The peak of the $90^{\circ}$ 
singularity starts to diminish and disappears quickly by cooling below 
19 K. In Fig. \ref{fig4} we exhibit two central line spectra one 
measured at 6 K and the other one at 294 K that have their intensities scaled 
to equal height for easier comparison. The peak of the $90^{\circ}$ 
singularity seen in the high temperature spectrum obviously is 
missing in the 6 K spectrum.  We explain this by the appearance of superconductivity in the 
grains with their c-axis close to the direction perpendicular to the 
external magnetic field.  Nevertheless, a substantial part of the 
spectrum coming from crystallites with a smaller $\theta$ remains 
unchanged down to 6 K. This indicates that in a magnetic field of 9 T at 
6 K a substantial part of the powder sample still remains normal 
conducting.  In view of the large $H_{c2}$ anisotropy quoted in the 
literature (2 -- 9) \cite{LimaPatnaikSimon} this comes not as a surprise.  
However, this has consequences for the determination of NMR quantities and their 
temperature dependence in the superconducting phase.  Necessarily one 
has to take into account the $H_{c2}$ anisotropy and avoid to average 
over the NMR signals coming from crystallites with different 
orientations in the magnetic field.  For reliable NMR measurements in 
the superconducting state large enough single crystals or aligned 
crystallites of $\mathrm{MgB_2}$ are most desirable.

The measurements of the NMR quantity $T_1$ were done by the
method of selective inversion of the central line and a subsequent
monitoring of the nuclear magnetization recovery $M(t)$ at
variable delay times t. The relaxation rate defined as
$\mathrm{1/T_1 = 2W}$, where W represents the spectral density of
the fluctuating internal magnetic fields, was extracted by fitting
the data to the recovery law
\begin{displaymath}
   \frac{ M(t)}{M(\infty)} =
  1 - \Gamma\left[{\textstyle \frac{1}{35}}\,e^{{\displaystyle -2Wt}} 
  +{\textstyle \frac{8}{45}}\,e^{{\displaystyle -12Wt}} 
  + {\textstyle \frac{50}{63}}\,e^{{\displaystyle -30Wt}}\right],
\end{displaymath}
obtained from the solution of the master equation for a spin 5/2
system in case of magnetic relaxation and selective excitation of
the central line \cite{AndrewSuter}.  The constant $\Gamma \leq
2$, also a fit parameter, depends on the inversion ratio of the
central line.  We measured $T_1$ only in the normal conducting
phase at three temperatures 30, 80 and 294 K.  The results are
presented in Table \ref{T1Tabelle}.  As expected for a metal the
product $T_1T$ remains constant.  The weighted average of the
three $T_1T$ values is $T_1T = 1090(50)$ sK.
\begin{table}[h]    
	\centering
	\caption{Temperature dependence of $^{25}$Mg T$_{1}$ and T$_{1}$T 
	in normal conducting $\mathrm{MgB_2}$} 
	\begin{ruledtabular}
	\begin{tabular}{ccc}   
		T(K) & T$_{1}$(s) & T$_{1}$T(sK)  \\
		\hline
		30 & 35(1.6) & 1050(50)  \\
		80 & 13.5(6) & 1080(50)  \\
		294 & 4.35(34) & 1280(100)  \\
	\end{tabular}
	\end{ruledtabular}
	\label{T1Tabelle}
\end{table}

What is the dominant mechanism of magnetic relaxation and Knight
shift of Mg and B in $\mathrm{MgB_2}$?  In most metals it is the
Fermi-contact interaction of the nucleus with the s-electrons at
the Fermi level. However, in case of $\mathrm{MgB_2}$ the states
near the Fermi level are mainly boron p-electron states with a
very small contribution of the s-electrons.  Pavarini {\em et
al.} \cite{Pavarini} making {\em ab initio} LDA calculations of
1/$T_1$ and $K$ at $\mathrm{^{11}B}$ and $\mathrm{^{25}Mg}$
sites in $\mathrm{MgB_2}$ show that the dominant relaxation
mechanism at the $\mathrm{^{11}B}$ nucleus is the interaction
with the electronic orbital moment.  For the $\mathrm{^{25}Mg}$
nucleus, however, they predict that the dominant relaxation
mechanism is the Fermi-contact interaction, which also dominates
the Mg Knight shift. They get for $\mathrm{^{25}Mg}$ $T_1T =
1000$ sK, $K_c = 256$ ppm, and $K_{ab} = 271$ ppm which is in
excellent agreement with our experimental values $T_1T = 1090(50)$
sK, $K_c = 242(4)$ ppm, and $K_{ab} = 252(6)$ ppm.  Forming the
Korringa ratio \cite{Carter} we get experimentally $(K^2T_1T)/S
\approx 0.95$, where $S =(\gamma_e/\gamma_n)^2 (h/8\pi)^2k_B =
7.0323\times10^{-5}$ sK for $\mathrm{^{25}Mg}$, with $\gamma_e$ and $\gamma_n$ the
gyromagnetic ratios for electron and nucleus, respectively.  The
observed experimental Korringa ratio of 0.95 is very close to the
ideal value of unity for s-electrons. This result is another
confirmation that the Fermi-contact interaction is indeed the
dominant mechanism responsible for relaxation and Knight shift at
the Mg site.

As next we compare the experimentally determined major principal
component, $V_{zz}$, of the EFG tensor at the Mg site with
$V_{zz}$ calculated {\em ab initio} within the density functional
theory by Tsvyashchenko {\em et al.} \cite{Tsvyashchenko} using the
full-potential linearized augmented plane wave method.  From the
experimentally determined $\mathrm{^{25}Mg}$ $\nu_{Q}$ and by use
of the $\mathrm{^{25}Mg}$ quadrupole moment $Q$ we extract the
absolute value $|V_{zz}| = 3.07(5)\times 10^{20}$
V/m$^2$ . The quadrupole coupling to the EFG and therefore $\nu_{Q}$
do not depend on the sign of $V_{zz}$.  The gained experimental
$|V_{zz}|$ has then to be compared to the absolute value of the
calculated $V_{zz} = -3.2\times10^{20}$ V/m$^2$
\cite{Tsvyashchenko}. We note that the agreement of both absolute values 
is excellent. Tsvashchenko {\em et al.} \cite{Tsvyashchenko}
also calculated $V_{zz}$ at the boron site where they get
$V_{zz}(B) = 18.5\times10^{20}$ V/m$^2$ which again is in good
agreement with the experimental value $V_{zz}(B) =
17.0(1)\times10^{20}$ V/m$^2$ \cite{Gerashenko,Jung}.  The ability
to reproduce the experimental NMR quantities at both Mg and B
sites by LDA calculations certainly strengthens the confidence
into the LDA approach to calculate other $\mathrm{MgB_2}$
material parameters.

The authors acknowledge useful discussions with M. Angst. The work 
was supported in part by the Swiss National Science Foundation.

\end{document}